\input{aipcheck}
\documentclass[
  ,final            % use final for the camera ready runs
 ]
{aipproc}

\layoutstyle{8x11double}

\usepackage{bm}

\begin{document}

\title{Nonequilibrium mode-coupling theory for uniformly sheared underdamped systems}

%<Replace this text with PACS numbers; choose from this list:
%\texttt{http://www.aip..org/pacs/index.html}>

\author{Koshiro Suzuki}{
  address={Canon Inc., 30-2 Shimomaruko 3-chome, Ohta-ku, Tokyo
  146-8501, Japan}
}

\author{Hisao Hayakawa}{
  address={Yukawa Institute for Theoretical Physics, Kyoto University,
  Kitashirakawa Oiwake-cho, Kyoto 606-8502, Japan}
}

\begin{abstract}
Mode-coupling theory for uniformly sheared underdamped systems with
an isothermal condition is presented.
%
%The controversial points in the previous formulations are correctly
%resolved.
%%
As a result of the isothermal condition, it is shown that the shear
stress exhibits significant relaxation at the $\alpha$-relaxation regime
due to the cooling effect which is accompanied by the growth of the
current fluctuation.
This indicates that nonequilibrium underdamped MCT is not equivalent
to the corresponding overdamped MCT, even at long time-scales after
the early stage of the $\beta$-relaxation.
It is verified that the effect of correlations to the density-current
modes is, however, negligible in sheared thermostated underdamped
systems.

\end{abstract}

\maketitle

%%%%%%%%%%%%%%%%%%%%%%%%%%%%%%%%%%%%%%%%%%%%
%% MAINMATTER
%%%%%%%%%%%%%%%%%%%%%%%%%%%%%%%%%%%%%%%%%%%%

\section{Introduction}

\hspace{0.5em}
Glassy liquids are extensively studied for understanding the transition
from liquid to amorphous solid states.
Two typical theoretical approaches coexist in these studies; the
mode-coupling theory (MCT) \cite{G}, and the replica theory with
assumptions on the metastable states from the random-first-order
transition theory (RFOT) \cite{PZ2010}.
Though not yet complete, the relation between the above two approaches
has been revealed in some aspects \cite{PZ2010, Szamel2010}.
For instance, it has been shown that the equation for the non-ergodic
parameter (NEP) in MCT can be partly derived by the replica theory
\cite{PZ2010}.
However, since the replica theory is a static theory, it lacks genuinely
dynamical effects, and its application to the study of dynamics,
e.g. relaxation towards steady state, or response, is difficult.

MCT, which is a dynamical theory, also suffers from difficulties when
applied to glassy liquids. %\cite{Miyazaki2007}.
For instance, it predicts unphysical non-ergodic transition, which
cannot be observed in realistic thermal systems \cite{2step1, KA1994}.
%
%However, this problem is at least formally avoided in {\it sheared}
%systems, which is the subject of the present paper.
%
Another issue is the lack of a consistent perturbative field-theoretic
formulation, which is practically applicable to higher orders.
Although efforts have been made \cite{ABL2006, KK2008, NH2008, JW2011},
they are still not comparable to the conventional approach of projection
operators and the Mori-type equations \cite{Z}, which we adopt
in this work.

Glassy liquids can be classified into two categories; one is molecular
glasses, which is described by Newtonian dynamics, and the other is
colloidal glasses, which is described by Brownian dynamics, at the
microscopic level, respectively.
Historically, MCT for molecular glasses has been developed, and has
passed various tests concerning the two-step relaxation phenomena \cite{G}.
Then, it was shown theoretically that the MCT for colloidal glasses are
equivalent to that for the molecular glasses in the long-time limit
\cite{SL1991}.
That is, the height of the plateau of the density time-correlator at the
$\beta$-relaxation regime (NEP), and the $\alpha$-relaxation time, are
coincident.
This fact has been verified by molecular dynamics (MD) simulations
\cite{GKB1998}.
Since then, it has been believed that the glassy behavior is insensitive
to its microscopic origin.

However, the established equivalency between the two categories is only
verified for {\it equilibrium} MCT, and is not yet established for {\it
nonequilibrium} MCT.
It is non-trivial whether nonequilibrium MCT for underdamped systems,
such as sheared molecular liquids or granular matters, share equivalent
long-time behaviors with that for overdamped systems.

Nonequilibrium MCT for uniformly sheared {\it overdamped} systems, for
e.g. colloidal suspensions immersed in solvents, has been worked out by
the pioneering papers of Fuchs and Cates \cite{FC2002, FC2009} and
Miyazaki {\it et al.} \cite{MR2002, MRY2004}.
In these works, it has been shown that the shear thinning and the
emergence of the yield stress in glassy states can be captured within
this framework.
Furthermore, comparison of MCT with MD has been performed in
Ref.~\cite{MRY2004}, and it has been shown that, aside from the
difference in the early stage before the $\beta$-relaxation due to
inertia effects, the plateau of the density time-correlator coincide
with each other.

On the other hand, nonequilibrium MCT for uniformly sheared {\it underdamped}
systems has been formulated by Chong and Kim \cite{CK2009}, whose
microscopic dynamics is described by the SLLOD equations \cite{SLLOD}.
%
%Besides the difference in the microscopic dynamics, apparent
%discrepancies between the resulting equations of Refs.~\cite{MRY2004,
%FC2009} and Ref.~\cite{CK2009} have been noticed.
%
However, there were some problems in this formulation, as indicated in
Ref.~\cite{SH2012}, and calculations to verify its validity have not yet
been performed.
In contrast to equilibrium MCT and sheared overdamped MCT, where the
temperature is a mere parameter, the temperature is a dynamic variable
in sheared underdamped MCT.
Hence, it is necessary to control the temperature to avoid heat-up or
cool-down in order to conform with realistic physical situations.
One of the problems of Ref.~\cite{CK2009} is that the temperature is not
controlled and heats up.

In this paper, we formulate the MCT for uniformly sheared {\it
underdamped} systems with an isothermal condition imposed.
%
%In a recent paper, we have reformulated the MCT for uniformly sheared
%{\it underdamped} systems to resolve the contradictions mentioned above
%\cite{SH2012}.
%%
%We have pointed out that the definition of the time-correlators in
%Ref.~\cite{CK2009} should be modified to satisfy the translational
%invariance in the sheared frame.
%%
%Our framework appears to be a non-trivial extension of
%Ref.~\cite{FC2009} to underdamped systems.
%
We present that there is a significant relaxation in the shear stress at
the $\alpha$-relaxation regime, although the profile of the density
time-correlator is coincident with the overdamped systems after the
early stage of the $\beta$-relaxation.
Then, we address another issue specific to underdamped systems: the
significance of correlations to the density-current modes.
We show that they are negligible in sheared underdamped thermostated
systems, but discuss that this situation might be accidental.
This result is new, which has not been discussed in Ref.~\cite{SH2012}.

%%%%%%%%%
% added
%%%%%%%%%
We stress that this paper is related to our previous paper \cite{SH2012}
in some respects, but it addresses issues not included there, such as
the effect of the density-current correlation, as well.
We also point out that our formulation of the underdamped MCT developed
in this paper can resolve the problem of the stress overshoot in the
$\alpha$-relaxation regime \cite{ZH2008, AW2013}.
It is shown that the overdamped MCT underestimates the stress overshoot
compared to MD \cite{ZH2008}, while it is implied that it is consistent
with the Brownian dynamics simulations \cite{AW2013}.
Thus, it might well be concluded that the pronounced stress overshoot in
molecular liquids than in colloidal suspensions is the result of the
inertia effect.

\section{Underdamped isothermal MCT}

\begin{figure}[tbh]
\includegraphics[width=8.5cm]{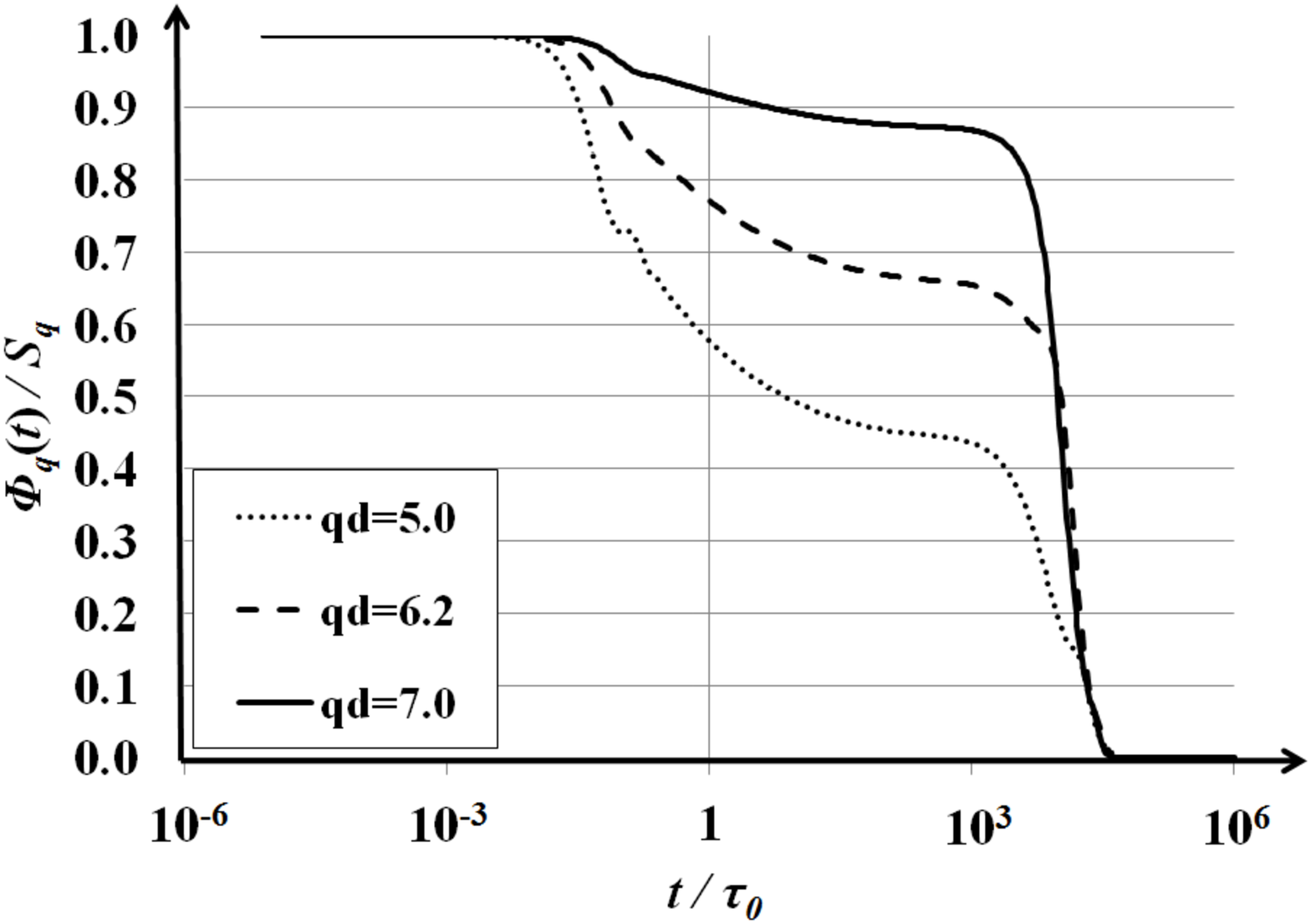}
\includegraphics[width=8.5cm]{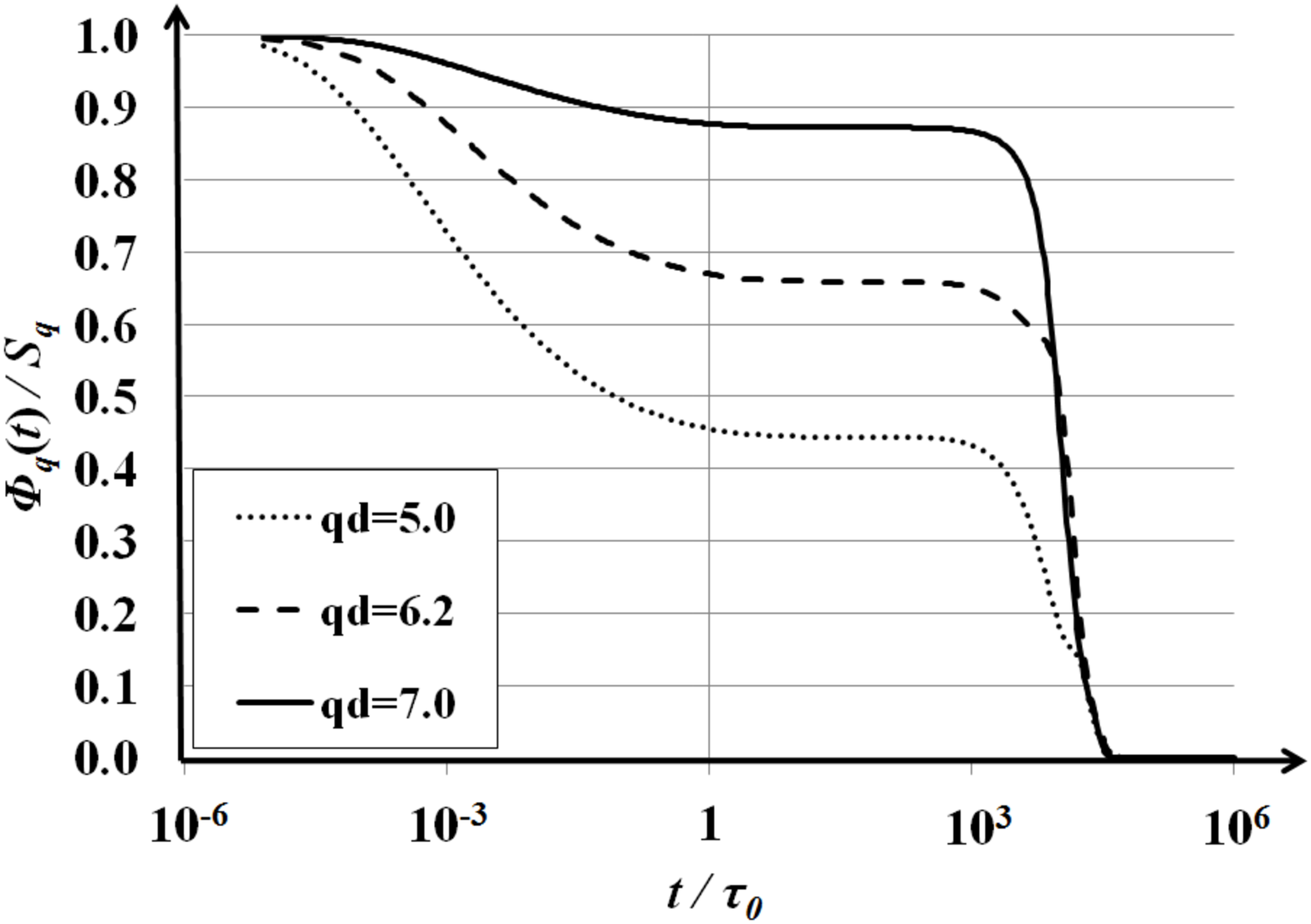}
\caption{
Time-evolution of the normalized density time-correlator
 $\Phi_q(t)/S_q$.
The left (right) figure is for the underdamped (overdamped) systems.
Time is scaled in unit of $\tau_0 \equiv d/v_0$,
 where $d$ is the diameter of the spherical particle and $v_0$ is the
 relative shear velocity at the boundaries.
The three lines correspond to nondimensionalized wavenumbers $qd =$ 5.0,
 6.2, 7.0.
The shear rate $\dot{\gamma}$ is chosen as $\dot{\gamma} \tau_0 =
 10^{-4}$, and the volume fraction $\varphi$ is chosen as $\epsilon \equiv (\varphi -
 \varphi_c)/\varphi_c = +10^{-3}$, where $\varphi_c \simeq 0.516$ is the
 critical volume fraction of the equilibrium MCT transition.}
\label{Fig:SH2012}
\end{figure}

\hspace{0.5em}
In this section, we present the MCT of uniformly sheared underdamped
systems, with isothermal condition imposed.
The basis of the formulation is the microscopic SLLOD equations
\cite{SLLOD} for a spherical $N$-body system of mass $m$,
\begin{eqnarray}
\dot{\bm{r}}_i(t)
&=&
\frac{\bm{p}_i(t)}{m} + \bm{\kappa}\cdot\bm{r}_i(t),
\\
\dot{\bm{p}}_i(t)
&=&
\bm{F}_i(t) - \bm{\kappa}\cdot\bm{p}_i(t) - \alpha(\bm{\Gamma}) \bm{p}_i(t),
\label{Eq:SLLOD_p}
\end{eqnarray}
where $i=1,\cdots,N$, and their corresponding Liouville equations.
Here, $\kappa^{\lambda \mu} = \dot{\gamma} \delta^{\lambda x}
\delta^{\mu y}$ ($\lambda, \mu = x,y,z$) is the shear-rate tensor, where
$\dot{\gamma}$ is the constant shear rate, and $\bm{F}_i(t)$ is a
conservative force with a soft-core potential whose details are
irrelevant in the following discussions.
The parameter $\alpha(\bm{\Gamma})$ in Eq.~(\ref{Eq:SLLOD_p}), where
$\bm{\Gamma} \equiv \{ \bm{r}_i, \bm{p}_i \}_{i=1}^N$ with $\bm{\Gamma}
\equiv \bm{\Gamma}(0)$, gives the dissipative coupling of the particle
with the thermostat.
If we impose the isothermal condition, which constrains the kinetic
temperature to remain constant from its initial equilibrium value,
$\alpha(\bm{\Gamma})$ should be regarded as a multiplier for this
constraint, rather than an independent physical parameter.
A representative example of this is the Gaussian isokinetic (GIK)
thermostat \cite{EM}, which is established as one of the most
typical thermostat in MD.
The specific form of $\alpha(\bm{\Gamma})$ in this thermostat is
\begin{eqnarray}
\alpha(\bm{\Gamma}) = 
\frac{\sum_{i=1}^N \left( \bm{F}_i \cdot \bm{p}_i -
\dot{\gamma} p_i^x p_i^y \right)}
{\sum_{j=1}^N \bm{p}_j \cdot
\bm{p}_j},
\label{Eq:GIK_alpha}
\end{eqnarray}
which follows from the constraint $\frac{d}{dt} \sum_{i=1}^N
\bm{p}_i(t)^2 = 0$.

Before the discussion of the isothermal condition in MCT, we address the
basics of the sheared underdamped MCT.
The major difference of the underdamped to the overdamped MCT is that,
besides the density time-correlator $\Phi_{\bm{q}}(t)$, we should
consider the current-density cross time-correlator
$H_{\bm{q}}^\lambda(t)$ ($\lambda = x,y,z$) on the same grounds.
A crucial feature of the sheared system is that translational invariance
is preserved in the sheared frame, although it is broken in the
experimental frame.
Hence, Fourier transform in the sheared frame is valid.
This leads to the following selection rule of wavevectors for two-point
functions, $\left\langle
A_{\bm{k}(t)}(\bm{\Gamma}(t))B_{\bm{q}}(\bm{\Gamma})^*
\right\rangle_{\mathrm{eq}} = \left\langle
A_{\bm{q}(t)}(\bm{\Gamma}(t))B_{\bm{q}}(\bm{\Gamma})^*
\right\rangle_{\mathrm{eq}} \delta_{\bm{k}, \bm{q}}$.
Here, $\bm{q}(t) \equiv \bm{q} - (\bm{q}\cdot\bm{\kappa})t$ is the
Affine-deformed wavevector, which incorporates the effect of shearing.
This selection rule leads to the following definitions of the above two
time-correlators,
\begin{eqnarray}
\Phi_{\bm{q}}(t) 
\hspace{-0.5em}
&\equiv&
\hspace{-0.5em}
\frac{1}{N}
\left\langle n_{\bm{q}(t)}(t)
n_{\bm{q}}(0)^* \right\rangle_{\mathrm{eq}},
\label{Eq:Phi}
\\
H_{\bm{q}}^\lambda(t) 
\hspace{-0.5em}
&\equiv&
\hspace{-0.5em}
\frac{i}{N} 
\left\langle j_{\bm{q}(t)}^\lambda(t)
n_{\bm{q}}(0)^* \right\rangle_{\mathrm{eq}},
\label{Eq:H}
\end{eqnarray}
where $n_{\bm{q}}(t)
\equiv \sum_{i=1}^N e^{i\bm{q}\cdot\bm{r}_i(t)} - N \delta_{\bm{q}, 0}$
and $j_{\bm{q}}^\lambda(t) \equiv \sum_{i=1}^N
p_i^\lambda(t)e^{i\bm{q}\cdot\bm{r}_i(t)}/m$ are the density and the
current-density fluctuations, respectively.
These definitions are not equivalent to those of Ref.~\cite{CK2009},
where Fourier transform is performed in the experimental frame.
%due to the non-Hermiticity and the non-commutativity of the shear part
%of the Liouvillian, $i\mathcal{L}_{\dot{\gamma}_r} \equiv \sum_{i=1}^N
%\bm{r}_i \cdot \bm{\kappa}^T \cdot (\partial / \partial \bm{r}_i)$.

Now we discuss the implementation of the isothermal condition.
The most straightforward and rigorous way to perform this is to adopt as
$\alpha(\bm{\Gamma})$ the one for the GIK thermostat,
Eq.~(\ref{Eq:GIK_alpha}).
If this is possible, the resulting Liouville and Mori-type equations are
expected to automatically satisfy the isothermal condition.
To rephrase it, not only the {\it averaged} temperature, but also the
{\it microscopic} temperature, is expected to remain constant.
However, this does not fit MCT, because the integral of a rational
function of momentum, Eq.~(\ref{Eq:GIK_alpha}), with Gaussian weight is
difficult to perform explicitly.
Hence, we propose another way, which only requires the {\it averaged}
temperature to remain constant.
This is attained by requiring the isothermal condition
to be satisfied at the level of the Mori-type equation, which is an
equation for the averaged time-correlator.
%, by the
%introduction of a multiplier.

The averaged temperature is defined by $\left\langle T(t)
\right\rangle_{\mathrm{eq}} \equiv 2 \left\langle K(\bm{\Gamma}(t))
\right\rangle_{\mathrm{eq}} / (3Nk_B)$, where $K(\bm{\Gamma}) \equiv
\sum_{i=1}^N \bm{p}_i^2 / (2m)$ is the total kinetic energy, and
$\left\langle \cdots \right\rangle_{\mathrm{eq}}$ represents the
equilibrium ensemble average.
The time derivative of the averaged temperature can be obtained from the
generalized Green-Kubo formula \cite{ME1987}, and the isothermal
condition reads
\begin{eqnarray}
\frac{d}{dt}
\left\langle
T(t)
\right\rangle_{\mathrm{eq}}
=
\frac{2}{3Nk_B}
\left\langle
K(\bm{\Gamma}(t)) \Omega(\bm{\Gamma}(0))
\right\rangle_{\mathrm{eq}}
= 0,
\label{Eq:GGK_T}
\end{eqnarray}
where  
\begin{eqnarray}
\Omega(\bm{\Gamma})
\equiv
- \beta V \dot{\gamma} \sigma_{xy}(\bm{\Gamma}) 
- 2 \beta \alpha(\bm{\Gamma}) \delta K(\bm{\Gamma})
\label{Eq:Omega}
\end{eqnarray}
is the work function.
Here, $V$ is the volume of the system, $\beta \equiv 1/(k_B
T_{\mathrm{eq}})$ is the inverse equilibrium temperature, $\delta
K(\bm{\Gamma})$ is the fluctuation of the kinetic energy, and
$\alpha(\bm{\Gamma})$ is a multiplier for the constraint
Eq.~(\ref{Eq:GGK_T}) which appears in the Liouvillian,
whose explicit form is discussed below.
The work function $\Omega(\bm{\Gamma})$ represents the gross work loaded
on the system by shearing and the dissipative coupling to the
thermostat.

By introducing general time-correlators of the form
\begin{eqnarray}
G_{A,B}(t)
\equiv
\left\langle
A(\bm{\Gamma}(t)) B(\bm{\Gamma})
\right\rangle_{\mathrm{eq}}
=
\left\langle
\left[
U(t) A(\bm{\Gamma}) 
\right]
B(\bm{\Gamma})
\right\rangle_{\mathrm{eq}}
,
\end{eqnarray}
where $U(t)$ is the time-evolution operator, the isothermal condition
Eq.~(\ref{Eq:GGK_T}) is written in terms of these time-correlators as
\begin{eqnarray}
\dot{\gamma} G_{K, \sigma}(t) +  2 G_{K, \alpha \delta K}(t) = 0.
\label{Eq:GGK_T2}
\end{eqnarray}
This condition indicates a balance between the average work by shearing
and the average dissipation.
We introduce a multiplier for the constraint Eq.~(\ref{Eq:GGK_T2}) in
$G_{K, \alpha \delta K}(t)$, which we denote by $\lambda_\alpha(t)$.
This is attained by defining a ``renormalized'' time-evolution operator
$U_R(t) \equiv \lambda_\alpha(t) U(t)$, and a corresponding
time-correlator $G_{K, \alpha \delta K}^{(\lambda)}(t)$, where
\begin{eqnarray}
G_{K, \alpha \delta K}^{(\lambda)}(t) 
&\equiv&
\hspace{-0.5em}
\left\langle
\left[
U_R(t) K(\bm{\Gamma}) 
\right]
\alpha(\bm{\Gamma}) \delta K(\bm{\Gamma})
\right\rangle_{\mathrm{eq}}
\nonumber \\
\hspace{-0.5em}
&=&
\hspace{-0.5em}
\lambda_\alpha(t)
G_{K, \alpha \delta K}(t),
\label{Eq:GKadK_lambda}
\end{eqnarray}
with which we recast Eq.~(\ref{Eq:GGK_T2}) as
\begin{eqnarray}
\dot{\gamma} G_{K, \sigma}(t) +  2 G_{K, \alpha \delta K}^{(\lambda)}(t) = 0.
\label{Eq:GGK_T3}
\end{eqnarray}
Note that $\lambda_\alpha(t)$ is a multiplier which is averaged and
appears in the Mori-type equations, and hence independent of
$\bm{\Gamma}$.
From Eqs.~(\ref{Eq:GKadK_lambda}) and (\ref{Eq:GGK_T3}), we can see that
the isothermal condition is satisfied by choosing $\lambda_\alpha(t)$ as
\begin{equation}
\lambda_\alpha(t)
=
- \frac{\dot{\gamma}}{2}
\frac{G_{K,\sigma}(t)}{G_{K, \alpha \delta K}(t)},
\label{Eq:isothermal}
\end{equation}
if $G_{K, \alpha \delta K}(t) \neq 0$.
On the other hand, if $G_{K, \alpha \delta K}(t) = 0$, the system lacks
cooling, and it inevitably heats up.

The Mori-type equations for the correlators $\Phi_{\bm{q}}(t)$ and
$H^\lambda_{\bm{q}}(t)$ are derived by introducing the projection
operator,
\begin{eqnarray}
\mathcal{P}(t) X
=
\sum_{\bm{k}}
\frac{\left\langle X n_{\bm{k}(t)}^* \right\rangle_{\mathrm{eq}}}{NS_{k(t)}}
n_{\bm{k}(t)}
+
\sum_{\bm{k}}
\frac{\left\langle X j_{\bm{k}(t)}^{\lambda *} \right\rangle_{\mathrm{eq}}}{Nv_T^2}
j_{\bm{k}(t)}^\lambda.
\end{eqnarray}
%
%and its complementary operator $\mathcal{Q}(t) \equiv 1 -
%\mathcal{P}(t)$.
%
Here, $S_k$ is the static structure factor and $v_T \equiv
\sqrt{T_{\mathrm{eq}}/m}$ is the thermal velocity, where
$T_{\mathrm{eq}}$ is the initial equilibrium kinetic temperature.
\begin{figure}[tbh]
\includegraphics[width=8.5cm]{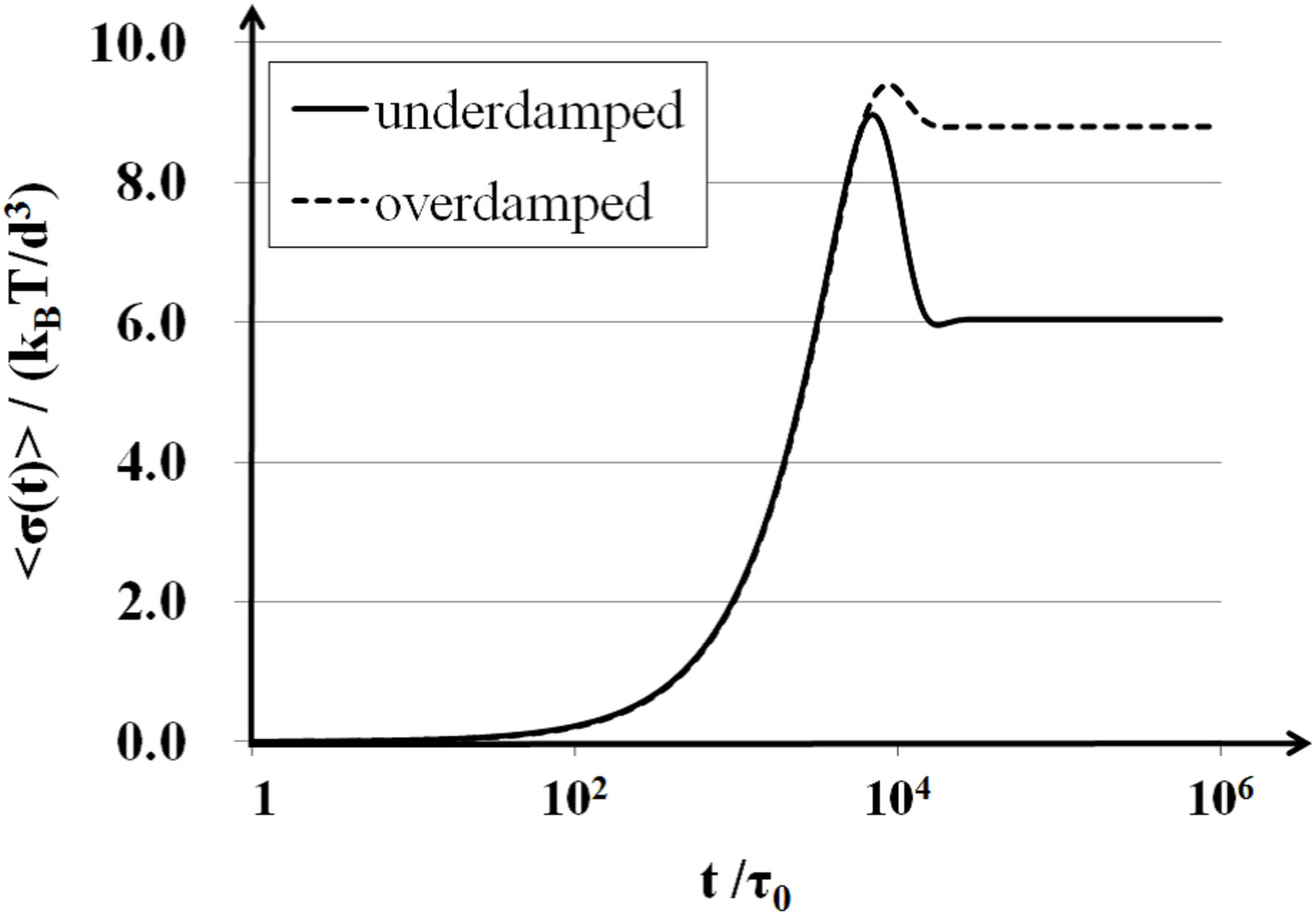}
\includegraphics[width=8.5cm]{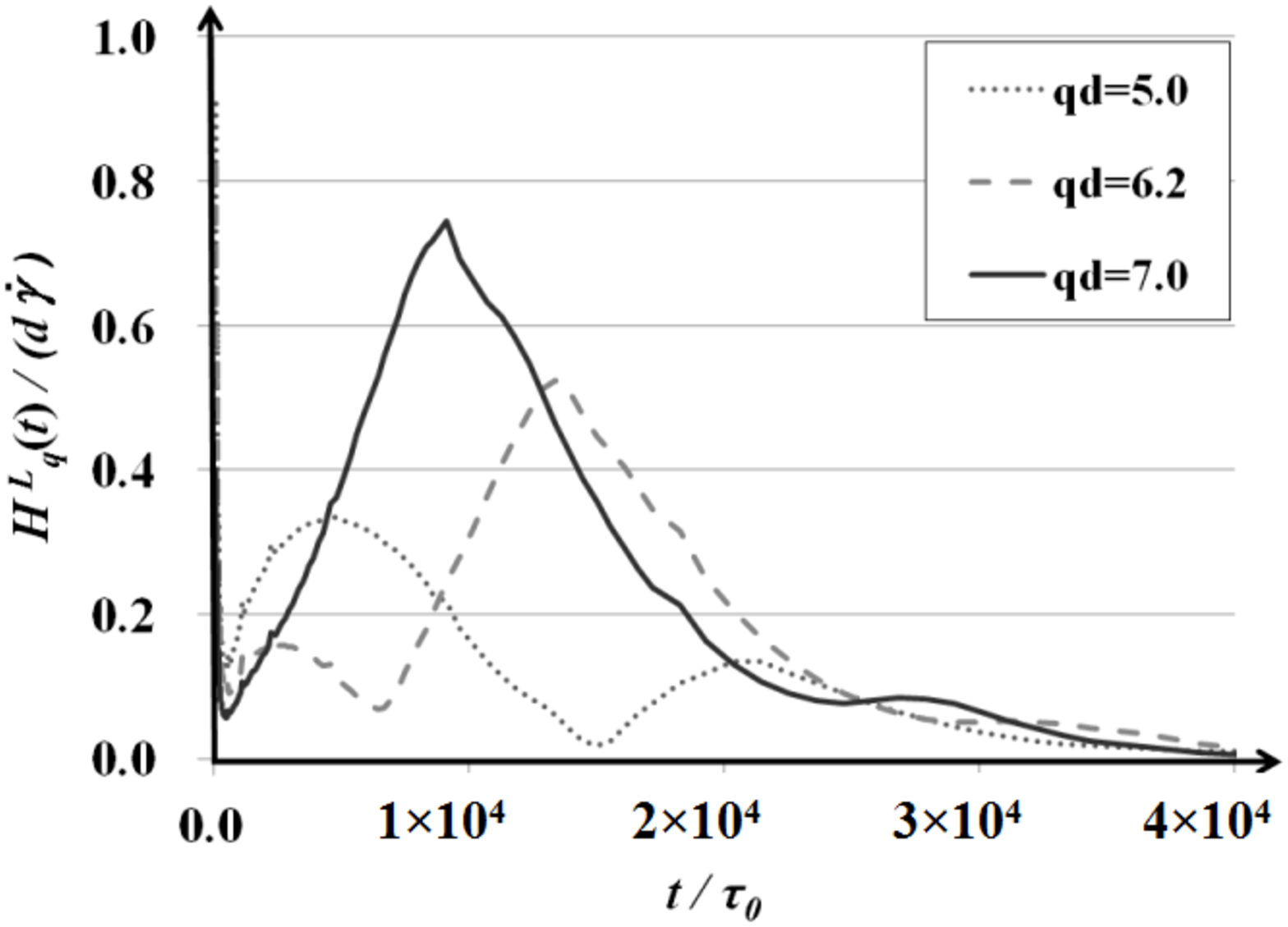}
 \caption{
Time-evolution of the average shear stress (left) and the longitudinal
 component of the current-density cross time-correlator $H^L_q(t)$ (right).
In the underdamped case, the shear stress is significantly relaxed at the
 $\alpha$-relaxation regime, which is accompanied by a growth of
 $H^L_q(t)$.
The shear rate is chosen as $\dot{\gamma} \tau_0 = 10^{-4}$, and the
 volume fraction as $\epsilon = (\varphi - \varphi_c)/\varphi_c = +10^{-3}$.}
\label{Fig:stress}
\end{figure}
To proceed further, it is necessary to resort to the mode-coupling
approximation, which determines the specific choice of
$\alpha(\bm{\Gamma})$ and $\lambda_\alpha(t)$, and provides a closure of
the Mori-type equations.
To this end, the second projecton operator to the pair-density modes
\begin{eqnarray}
\mathcal{P}_{nn}(t) X
=
\sum_{\bm{k}>\bm{p}} 
\frac{ \left\langle X n_{\bm{k}(t)}^* n_{\bm{p}(t)}^* \right\rangle_{\mathrm{eq}} }
{ N^2 S_{k(t)}S_{p(t)} }
n_{\bm{k}(t)} n_{\bm{p}(t)} 
\label{Eq:Pnn}
\end{eqnarray}
is introduced. 
The validity of this choice will be discussed in the next section.
The simplest form of $\alpha(\bm{\Gamma})$ is $\alpha(\bm{\Gamma}) =
\alpha_0 = {\rm const.}$, where no fluctuation is incorporated.
This is the choice adopted in Ref.~\cite{CK2009}.
However, we have figured out that this choice leads to $G_{K, \alpha
\delta K}(t) = 0$; the isothermal condition cannot be satisfied, and the
system heats up.
%
%Furthermore, since the ``cooling effect'' $G_{K, \alpha \delta K}^{(\lambda)}(t)$
%vanishes, the average kinetic temperature heats up, which converges into
%a value $\lim_{t\to \infty} \left\langle T(t)
%\right\rangle_{\mathrm{eq}}$ not affected by the value of $\alpha_0$.
%%
%It is obvious that this unphysical temperature cannot be regarded as a
%steady-state kinetic temperature.
%%
%
Hence, it is necessary to incorporate fluctuations into
$\alpha(\bm{\Gamma})$. 
In this respect, the simplest form would be
\begin{eqnarray}
\alpha(\bm{\Gamma})
=
\frac{\alpha_0}{\frac{3}{2}N k_B T_{\mathrm{eq}}}
\sum_i \frac{\bm{p}_i^2}{2m},
\label{Eq:alpha}
\end{eqnarray}
where current fluctuations are incorporated.
Here, $\alpha_0$ is a constant.
With this choice, it can be verified that $G_{K, \alpha \delta K}(t)
\neq 0$, and hence the constraint Eq.~(\ref{Eq:GGK_T3}) can be
satisfied.
Straightforward calculation of $G_{K, \alpha \delta K}(t)$ and $G_{K,
\sigma}(t)$ leads to
\begin{eqnarray}
\lambda_\alpha(t)
\alpha_0
=
\frac{\dot{\gamma}}{2}
\frac{
\sum_{\bm{k}>0}\frac{1}{S_{k(t)}S_k}
W_{\bm{k}}
\Phi_k(t)^2
}
{
\sum_{\bm{k}>0}\frac{1}{S_{k(t)}S_k}
\Phi_k(t)^2
},
\label{Eq:alpha0}
\end{eqnarray}
where $W_{\bm{k}} \equiv \frac{k^x k^y}{k} \frac{1}{S_k} \frac{\partial
S_k}{\partial k} $ \cite{SH2012}.
This expression corresponds to $\alpha(\bm{\Gamma})$ of the GIK
thermostat; it is a ratio of the work loaded on the system (numerator)
to the total kinetic energy (denominator).

Now we present the Mori-type equation with the mode-coupling
approximation applied.
We adopt the isotropic approximation, in which case the two Mori-type
equations can be cast in a single second-order equation for
$\Phi_{q}(t)$, 
\begin{eqnarray}
\frac{d^2}{dt^2} \Phi_q(t)
\hspace{-0.5em}
&=&
\hspace{-0.5em}
-
\left[
\lambda_\alpha(t) \alpha_0
-
\dot{\gamma}
\frac{\frac{2}{3}\dot{\gamma}t}{1 + \frac{1}{3}(\dot{\gamma}t)^2}
\right]
\frac{d}{dt}
\Phi_q(t)
\nonumber \\
&&
\hspace{-4em}
-
v_T^2 \frac{q(t)^2}{S_{q(t)}} 
\Phi_q(t)
-
\int_0^t ds
M_{q(s)}(t-s)
\frac{d}{ds} \
\Phi_q(s),
\hspace{2em}
\label{Eq:MCTeq}
\end{eqnarray}
where the multiplier $\lambda_\alpha(t)$ appears as an effective
friction coefficient.
It is notable that the memory kernel is unaltered by the introduction of
the multiplier with the choice of the second projection operator
Eq.~(\ref{Eq:Pnn}).
In the isotropic approximation, only the longitudinal component of the
current-density cross time-correlator, $H^L_q(t) = d \Phi_q(t)/dt$, is
considered.  
The numerical solution of Eq.~(\ref{Eq:MCTeq}) is shown in
Fig.~\ref{Fig:SH2012}, together with the corresponding result for the
overdamped case.
As can be seen, the long-time behavior after the early stage of the
$\beta$-relaxation, namely the height of the plateau of the density
time-correlator and the $\alpha$-relaxation time, is equivalent for the
two cases.
%
%there appear unphysical undershoots and overshoots in
%the result of Ref.~\cite{CK2009}, while these are absent in that of
%Ref.~\cite{SH2012}. 
%

Despite this fact, there is a significant difference between underdamped
and overdamped MCT for sheared systems.
From the generalized Green-Kubo formula, the following formula for the
shear stress can be derived for the underdamped case \cite{SH2012},
\begin{eqnarray}
&&
\hspace{-3em}
\left\langle
\sigma (t)
\right\rangle_{\mathrm{eq}}
=
\frac{k_B T_{\mathrm{eq}}}{2}
\int_0^t ds
\int \frac{d^3 \bm{k}}{(2\pi)^3}
\nonumber \\
&&
\hspace{2em}
\times
\frac{W_{\bm{k}(s)}}{S_{k(s)}}
\frac{\dot{\gamma}W_{\bm{k}}- 2\lambda_\alpha(s) \alpha_0}{S_k}
\Phi_{\bm{k}}(s)^2.
\label{Eq:stress}
\end{eqnarray}
The corresponding well-known formula for the overdamped case
\cite{FC2002} can be obtained by setting $\lambda_\alpha(t) \alpha_0 =
0$ in Eq.~(\ref{Eq:stress}).
This term with $\lambda_\alpha(t) \alpha_0$ is exactly the ``cooling
effect'' specific for sheared underdamped systems, which prevents the
system from heating up.
The numerical result of Eq.~(\ref{Eq:stress}), together with that of
$H^L_{q}(t)$, is exhibited in Fig.~\ref{Fig:stress}.
We can see that, in the underdamped case, there is a significant
relaxation of the shear stress in the $\alpha$-relaxation regime.
Furthermore, this relaxation is accompanied by a growth of $H^L_q(t)$.
From these results, we can obtain an intuitive picture as follows;
during the $\beta$-relaxation regime, the particles are caged by their
neighbors, and they are almost frozen.
When the accumulated strain by shearing becomes comparable to the
particle size, the cage is broken, and the partiles start to escape from
the cage.
This generates current fluctuations, which is cooled to keep the
temperature of the system unchanged.
This cooling effect leads to a relaxation of the system to the genuine
stable state, and a stress relaxation occurs.

It is remarkable that the overdamped MCT underestimates the stress
overshoot in the $\alpha$-relaxation regime compared to its
corresponding MD \cite{ZH2008}.
On the other hand, it is impled that the overdamped MCT predicts the
magnitude of the stress overshoot consistently with the Brownian
dynamics simulations \cite{AW2013}.
This consistency is established by means of the schematic model, rather
than the microscopic model, of the MCT, so a complete verification
remains to be a future problem.
However, we might say that the results of Refs.~\cite{ZH2008} and
\cite{AW2013} strongly suggest that the pronounced stress overshoot
in molecular liquids than in colloidal suspensions is the result of the
inertia effect.

\section{Effect of density-current modes}
\label{sec:Pnj}
\begin{figure}[thb]
\includegraphics[width=8.5cm]{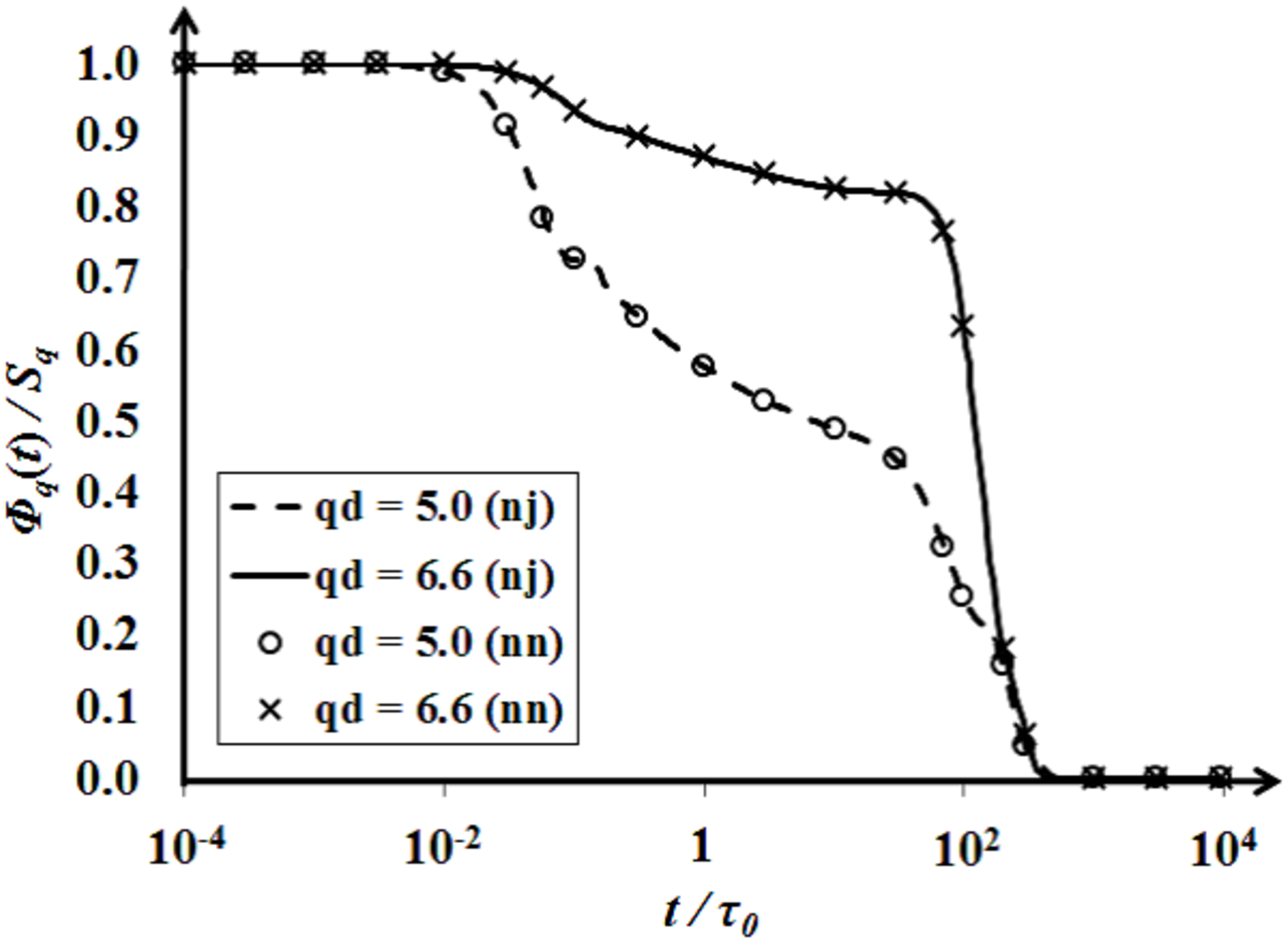}
\includegraphics[width=8.5cm]{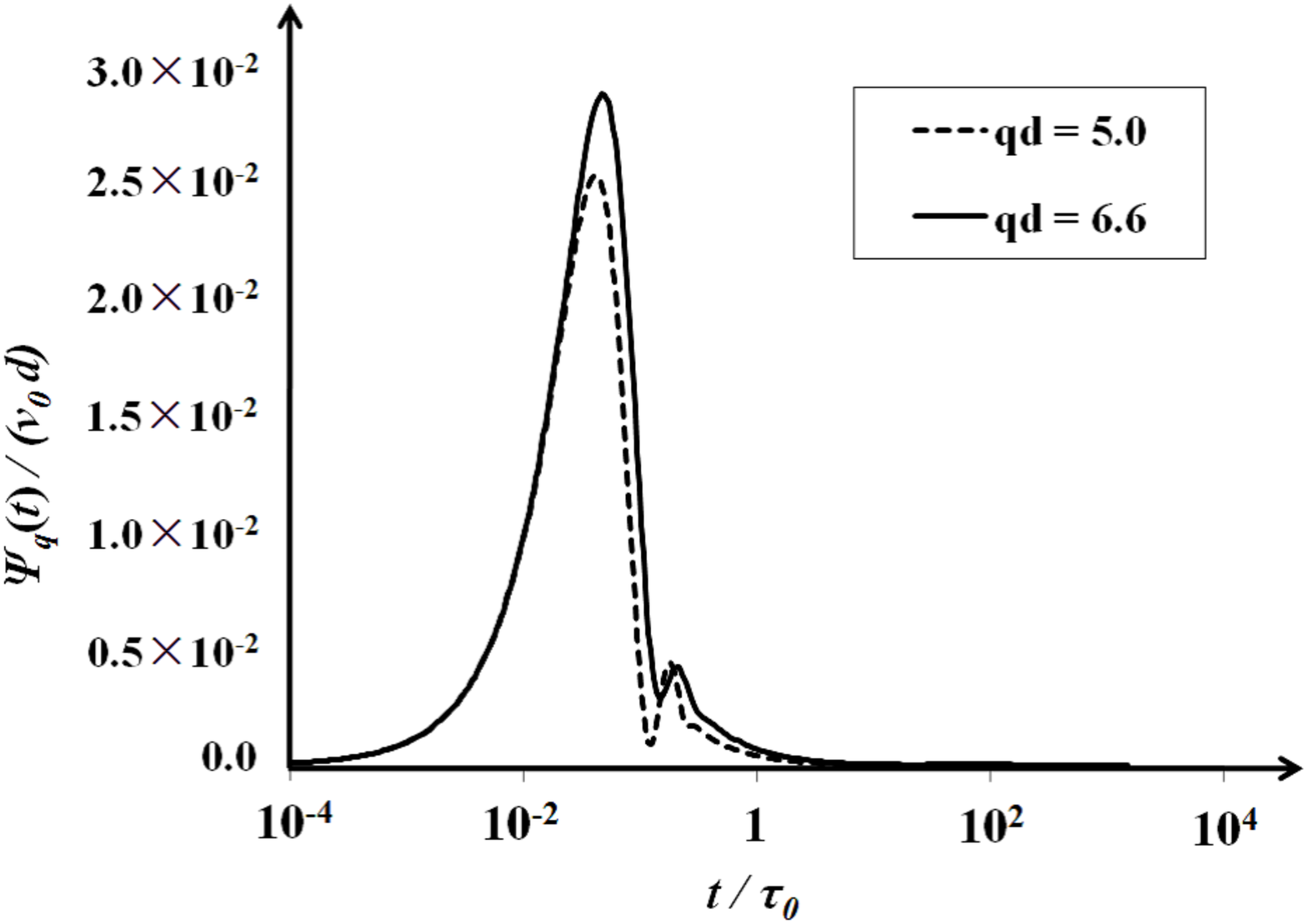}
\caption{Numerical results for the time-correlators $\Phi_q(t)$ (left)
and $\Psi_q(t)$ (right). In the left figure, solid and dashed lines
with caption (nj) are the solutions of Eq.~(\ref{Eq:Phi_eq_nj}) with $\mathcal{P}_2(t)=\mathcal{P}_{nn}(t)+\mathcal{P}_{nj}(t)$, 
while circles and crosses with caption (nn) are the solutions for $\mathcal{P}_2(t)=\mathcal{P}_{nn}(t)$.}
\label{Fig:PhiPsi_nj}
\end{figure}
%
%\end{widetext}

\hspace{0.5em} 
In the previous section, we have addressed a problem specific to the
sheared underamped MCT; that is, to control the temperature from
heating-up or cooling-down.
There is another non-trivial problem specific to the sheared underdamped
MCT which should be addressed; ``up to what degree should we take
correlations into account?''.
The fundamental concept of MCT is to take into account correlations with
``slow'' modes.
In the overdamped case, it is the density fluctuation $n_{\bm{q}}(t)$,
and in the underdamped case, it is $n_{\bm{q}}(t)$ and the
current-density fluctuation $j_{\bm{q}}^\lambda(t)$.
Although $j_{\bm{q}}^\lambda(t)$ is ``faster'' than $n_{\bm{q}}(t)$, it
is necessary to incorporate to deal with physical phenomena such as
stress relaxation at the $\alpha$-relaxation regime, which we have shown
in the previous section.
Incorporating correlations, or projecting onto $n_{\bm{q}}(t)$ and/or
$j_{\bm{q}}^\lambda(t)$, leads to the Mori-type equations, which
includes a memory kernel as a time-correlator of noises (uncorrelated
degrees of freedom).
A memory kernel includes higher-order correlations, and we should
extract these correlations by projecting onto higher-order modes, such
as $nn$, $nj$, $nnn$, $\cdots$.
(Of course, correlations with higher-order modes correspond to
multi-body correlations, which must be truncated to close the equation.
MCT is a scheme which approximates a four-body correlation by
factorizing it into a product of two-body correlations.)
For this purpose, we have introduced the second projection operator
Eq.~(\ref{Eq:Pnn}), which is conventional in equilibrium as well as
in sheared overdamped MCT.
However, in the sheared underdamped MCT, it is not {\it a priori}
apparent that the projection onto the density-current modes
\begin{eqnarray}
\mathcal{P}_{nj} (t) X
\hspace{-0.5em}
&\equiv&
\hspace{-0.5em}
\sum_{\bm{k} , \bm{p}}
\frac{
\left\langle X n_{\bm{k}(t)}^* j_{\bm{p}(t)}^{\lambda*} \right\rangle 
}{N^2 S_{k(t)} v_T^2}
n_{\bm{k}(t)} j_{\bm{p}(t)}^\lambda
\label{Eq:Pnj}
\end{eqnarray}
is negligible compared to $\mathcal{P}_{nn}(t)$, and hence the adoption
of Eq.~(\ref{Eq:Pnn}) must be validated.
In particular, we should carefully check the roles of ${\mathcal
P}_{nj}(t)$ as well as the self-consistency of our argument in the
previous section in our underamped MCT, because the existence of the
momentum current causes some essential differences in the long-time
dynamics from that in the overdamped MCT.
In this section, we discuss the effect of the projection onto the
density-current modes in the second projection operator,
Eq.~(\ref{Eq:Pnj}).

Let us consider a second projection operator $\mathcal{P}_2(t)$ of the
following form \cite{CSOH2012},
\begin{eqnarray}
\mathcal{P}_2 (t)
\hspace{-0.5em}
&=&
\hspace{-0.5em}
\mathcal{P}_{nn} (t)
+
\mathcal{P}_{nj} (t),
\label{Eq:Pnj}
\end{eqnarray}
where $\mathcal{P}_{nn}(t)$ and $\mathcal{P}_{nj}(t)$ are defined by
Eqs.(\ref{Eq:Pnn}) and (\ref{Eq:Pnj}), respectively.
We note that the isothermal condition Eq.~(\ref{Eq:GGK_T3}) and the
stress formula Eq.~(\ref{Eq:stress}) are unaltered by the introduction
of $\mathcal{P}_{nj}(t)$.
Crucial observation for this property is that the following equalities
hold, $\mathcal{P}_{nj}(t) \sigma_{xy}(\bm{\Gamma}) = 0$,
$\mathcal{P}_{nj}(t) K(\bm{\Gamma}) = 0$, and $\mathcal{P}_{nj}(t)
\left[ \alpha(\bm{\Gamma}) \delta K (\bm{\Gamma})\right] = 0$, which
follow from the vanishing of ensemble averages with odd number of
momentum variables.
Here, the explicit form Eq.~(\ref{Eq:alpha}) of $\alpha(\bm{\Gamma})$ is
assumed.
As a result, the time-correlators $G_{K, \sigma}(t)$, $G_{K, \alpha
\delta K}(t)$, $G_{\sigma, \sigma}(t)$, and $G_{\sigma, \alpha \delta
K}(t)$ remain unchanged.
The isothermal condition and the stress formula are expressed in terms
of $G_{K, \alpha \delta K}(t)$, $G_{\sigma, \alpha \delta K}(t)$ and
$G_{K, \sigma}(t)$, $G_{\sigma, \sigma}(t)$, respectively, so the above
statement is proved.

Corrections due to $\mathcal{P}_{nj}(t)$ arise in the Mori-type
equations for the time-correlators.
The introduction of $\mathcal{P}_{nj}(t)$ requires us to introduce two
additional time-correlators defined as follows,
\begin{eqnarray}
\bar{H}_{\bm{q}}^\lambda(t)
\hspace{-0.5em}
&\equiv&
\hspace{-0.5em}
\frac{i}{N}
\left\langle
n_{\bm{q}(t)}(t) j_{\bm{q}}^{\lambda *}(0)
\right\rangle,
\label{Eq:bH}
\\
C_{\bm{q}}^{\lambda\mu}(t)
\hspace{-0.5em}
&\equiv&
\hspace{-0.5em}
\frac{1}{N}
\left\langle
j_{\bm{q}(t)}^\lambda(t) j_{\bm{q}}^{\mu *}(0)
\right\rangle,
\label{Eq:C}
\end{eqnarray}
in addition to the previously introduced $\Phi_{\bm{q}}(t)$ and
$H_{\bm{q}}^\lambda(t)$, which are defined in Eqs.~(\ref{Eq:Phi}) and
(\ref{Eq:H}), respectively.
Note that $H_{\bm{q}}^\lambda(t)$ and $\bar{H}_{\bm{q}}^\lambda(t)$ are
inequivalent, since sheared systems do not possess translational and
reversal symmetries in time.
Similarly to $\Phi_{\bm{q}}(t)$ and $H_{\bm{q}}^\lambda(t)$, we adopt the
isotropic approximation for $\bar{H}_{\bm{q}}^\lambda(t)$ and
$C_{\bm{q}}^{\lambda\mu}(t)$.
By introducing two scalar time-correlators, $\Psi_q(t)$ and $C_q(t)$,
it reads
\begin{eqnarray}
\bar{H}_{\bm{q}}^\lambda(t)
&\simeq&
-q(t)^\lambda \Psi_q(t),
\\
C_{\bm{q}}^{\lambda\mu}(t)
&\simeq&
\delta^{\lambda\mu}
C_q(t).
\end{eqnarray}
By this approximation, it is possible to derive the following equations
for the time-correlators,
\begin{eqnarray}
\frac{d^2}{dt^2}
\Phi_q(t)
\hspace{-0.5em}
&=&
\hspace{-0.5em}
-
v_T^2
\frac{q(t)^2}{S_{q(t)}}
\Phi_q(t)
\nonumber \\
&&
-
\left[
\lambda_\alpha(t) \alpha_0
-
2
f_1\left( \dot{\gamma}, t \right)
\right]
\frac{d}{dt}
\Phi_q(t)
\nonumber \\
&&
-
\int_0^t ds
M_{\bm{q}(s)}(t-s)
\frac{d}{ds}
\Phi_{q}(s),
\label{Eq:Phi_eq_nj}
\end{eqnarray}
\begin{eqnarray}
\frac{d^2}{dt^2}
\Psi_q(t)
\hspace{-0.5em}
&=&
\hspace{-0.5em}
-
\left[
\frac{v_T^2}{3}
\frac{q(t)^2}{S_{q(t)}}
+
\lambda_\alpha(t) \alpha_0
f_1\left( \dot{\gamma}, t \right)
\right.
\nonumber \\
&&
\left.
+
f_2\left( \dot{\gamma}, t \right)
\right]
\Psi_q(t)
\nonumber \\
&&
-
\left[
\lambda_\alpha(t) \alpha_0
+
f_1\left( \dot{\gamma}, t \right)
\right]
\frac{d}{dt}
\Psi_q(t)
\nonumber \\
&&
-
\frac{1}{3}
\int_0^t ds
M_{\bm{q}(s)}^{\lambda \lambda}(t-s)
\nonumber \\
&&
\times
\left[
\frac{d}{ds} \Psi_q(s)
+
f_1\left( \dot{\gamma}, s \right)
\Psi_q(s)
\right],
\label{Eq:Psi_eq_nj}
\end{eqnarray}
where functions $f_1(\dot{\gamma}, t)$ and $f_2(\dot{\gamma}, t)$ which
incorporate anisotropic effects of shearing are defined as
\begin{eqnarray}
f_1\left( \dot{\gamma}, t \right)
&\equiv&
\frac{\frac{1}{3}
\dot{\gamma}^2
t}
{1 +
\frac{1}{3}
\dot{\gamma}^2 t^2},
\\
f_2\left( \dot{\gamma}, t \right)
&\equiv&
\frac{\frac{1}{3}
\dot{\gamma}^2
\left(
1 -
\frac{1}{3}
\dot{\gamma}^2 t^2
\right)
}
{
\left(
1 +
\frac{1}{3}
\dot{\gamma}^2 t^2
\right)^2}.
\end{eqnarray}
Here, we have eliminated $C_q(t)$ by the relation
\begin{eqnarray}
C_q(t)
&=&
\frac{d}{dt}
\Psi_q(t)
+
\frac{1}{2}
\frac{\Psi_q(t)}{q(t)^2}
\frac{d}{dt}
q(t)^2,
\end{eqnarray}
which follows from the Mori-type equation.
The memory kernel in Eq.~(\ref{Eq:Phi_eq_nj}) consists of two terms,
which are given by
%
%\begin{widetext}
\begin{eqnarray}
M_{\bm{q}}(\tau)
\hspace{-0.5em}
&\equiv&
\hspace{-0.5em}
\sum_{i=1,2}
M_{\bm{q}}^{(i)}(\tau),
\nonumber \\
M_{\bm{q}}^{(1)}(\tau)
\hspace{-0.5em}
&=&
\hspace{-0.5em}
\frac{nv_T^2}{2q^2}
\left[
1
+
\frac{1}{3}
\left( \dot{\gamma} \tau \right)^2
\right]
\int \frac{d^3 \bm{k}}{(2\pi)^3}
\nonumber \\
&&
\hspace{-3em}
\times
\left[
(\bm{q}\cdot\bm{k}) c_{\bar{k}(\tau)}
+
(\bm{q}\cdot\bm{p}) c_{\bar{p}(\tau)}
\right]
\left[
(\bm{q}\cdot\bm{k}) c_k
+
(\bm{q}\cdot\bm{p}) c_p
\right]
\nonumber \\
&&
\hspace{-3em}
\times
\Phi_k(\tau)
\Phi_p(\tau),
\\
%%%
M_{\bm{q}}^{(2)}(\tau)
\hspace{-0.5em}
&=&
\hspace{-0.5em}
-
\frac{2\lambda_\alpha(\tau) \alpha_0}{q^2}
\left[
1
+
\frac{1}{3}
\left( \dot{\gamma} \tau \right)^2
\right]
\int\frac{d^3 \bm{k}}{(2\pi)^3}
\nonumber \\
&&
\hspace{-2em}
\times
\left[
(\bm{q}\cdot\bm{k}) c_{\bar{k}(\tau)}
+
(\bm{q}\cdot\bm{p}) c_{\bar{p}(\tau)}
\right]
\left(
\bm{q} \cdot \bm{p}
\right)
\nonumber \\
&&
\hspace{-2em}
\times
\Phi_k(\tau)
\Psi_p(\tau).
\label{Eq:M2}
\end{eqnarray}
%\end{widetext}
%
Note that the first term $M_{\bm{q}}^{(1)}(\tau)$ is nothing but the
conventional memoery kernel familiar in equilibrium as well as in
sheared overdamped MCT, which is the source of the plateau in the
density time-correlator.
The second term $M_{\bm{q}}^{(2)}(\tau)$ is the novel therm which
originates in the ``cooling effect'' due to the isothermal condition.
The memory kernel in Eq.~(\ref{Eq:Psi_eq_nj}), which is the trace of
$M_{\bm{q}}^{\lambda\mu}(\tau)$, also consists of two terms, which are
given by
%
%\begin{widetext}
\begin{eqnarray}
M_{\bm{q}}^{\lambda \lambda}(\tau)
\hspace{-0.5em}
&=&
\hspace{-0.5em}
\sum_{i=1,2}
M_{\bm{q}}^{(i)\lambda \lambda}(\tau),
\nonumber \\
M_{\bm{q}}^{(1)\lambda \lambda}(\tau)
\hspace{-0.5em}
&=&
\hspace{-0.5em}
\frac{nv_T^2}{2}
\int \frac{d^3 \bm{k}}{(2\pi)^3}
\left[
\bm{k} c_k
+
\bm{p} c_p
\right]
\cdot
\left[
\bm{k} c_{\bar{k}(\tau)}
+
\bm{p} c_{\bar{p}(\tau)}
\right]
\nonumber \\
&&
\times
\Phi_k(\tau)
\Phi_p(\tau),
\\
%%%
M_{\bm{q}}^{(2)\lambda \lambda}(\tau)
\hspace{-0.5em}
&=&
\hspace{-0.5em}
-
2 \lambda_\alpha(\tau) \alpha_0
\left[
1 +
\frac{1}{3}
\left( \dot{\gamma} t \right)^2
\right]
\int\frac{d^3 \bm{k}}{(2\pi)^3}
\nonumber \\
&&
\times
\bm{p}
\cdot
\left[
\bm{k} c_{\bar{k}(\tau)}
+
\bm{p} c_{\bar{p}(\tau)}
\right]
\Phi_k(\tau)
\Psi_p(\tau),
\label{Eq:TrM2}
\end{eqnarray}
%\end{widetext}
%
where again the second term originates in the ``cooling effect''.
The two correlators $\Phi_q(t)$ and $\Psi_q(t)$ couple through
$M_{\bm{q}}^{(2)}(\tau)$ and $M_{\bm{q}}^{(2)\lambda\lambda}(\tau)$.
The derivation of Eqs.~(\ref{Eq:Phi_eq_nj})--(\ref{Eq:TrM2}) is
somewhat lengthy, though straightforward, so we present only the
results.
%, and the details will be reported elsewhere \cite{CSOH2012}.

It should be noted that, in the absense of the coupling to the
thermostat, $M_{\bm{q}}^{(2)}(\tau)$ and
$M_{\bm{q}}^{(2)\lambda\lambda}(\tau)$ vanish.
This suggests that the memory kernel is coincident for the equlibrium
underdamped and overdamped MCT, even when projection onto the
density-current modes $\mathcal{P}_{nj}(t)$ is included.
This proves the validation of the projection onto the pair-density modes
$\mathcal{P}_{nn}(t)$ in the equilibrium underdamped MCT.

The numerical solutions of Eqs.~(\ref{Eq:Phi_eq_nj}) and
(\ref{Eq:Psi_eq_nj}) are shown in Fig.~\ref{Fig:PhiPsi_nj}.
For the density time-correlator $\Phi_q(t)$, the results for the case
without including $\mathcal{P}_{nj}(t)$, i.e. $\mathcal{P}_2(t) =
\mathcal{P}_{nn}(t)$, are plotted in circles and crosses.
The initial conditions are chosen as follows:
\begin{eqnarray}
\Phi_q(t=0)
&=&
S_q,
\hspace{1em}
\left[
\frac{d}{dt}
\Phi_q(t)
\right]_{t=0}
=
0,
\\
\Psi_q(t=0)
&=&
0,
\hspace{1em}
\left[
\frac{d}{dt}
\Psi_q(t)
\right]_{t=0}
=
\frac{k_B T}{m}.
\end{eqnarray}
As can be seen from Fig.~\ref{Fig:PhiPsi_nj}, although a sign of
correlation of currents can be seen in $\Psi_q(t)$ (which appears as a
peak around $t/\tau_0 \simeq 10^{-1}$), the result of $\Phi_q(t)$ is
almost coincident with that for $\mathcal{P}_2(t) =
\mathcal{P}_{nn}(t)$.
Together with the fact that $\mathcal{P}_2(t)$ of Eq.~(\ref{Eq:Pnj})
leads to the identical isothermal condition and the stress formula as
for $\mathcal{P}_2(t) = \mathcal{P}_{nn}(t)$, it is verified that
projection onto the density-current modes is negligible in sheared
thermostated systems as well.
Thus, the analysis in the previous section, where we only use ${\mathcal
P}_{nn}(t)$ for the second projection operator, is self-consistent.

In principle, we can see from Eqs.~(\ref{Eq:M2}) and (\ref{Eq:TrM2})
that the ``cooling effect'' which affects the memory kernel works to
destroy the plateau of the density time-correlator, since their
contributions are negative.
This can be interpreted as the exhausion of the cage by cooling, which
allows the caged particles to escape.
However, as shown in Fig.~\ref{Fig:stress}, $M_{\bm{q}}^{(2)}(\tau)$ and
$M_{\bm{q}}^{(2)\lambda\lambda}(\tau)$ becomes significant only at the
$\alpha$-relaxation regime, where the plateau-generating kernels
$M_{\bm{q}}^{(1)}(\tau)$ and $M_{\bm{q}}^{(1)\lambda\lambda}(\tau)$
decay irrespective of the existence of $M_{\bm{q}}^{(2)}(\tau)$ and
$M_{\bm{q}}^{(2)\lambda\lambda}(\tau)$.
This is why the effect of $M_{\bm{q}}^{(2)}(\tau)$ and
$M_{\bm{q}}^{(2)\lambda\lambda}(\tau)$ is hidden behind.

Although the role of $\mathcal{P}_{nj}(t)$ is not significant in the
present case, it is an important feature that dissipation acts to
destroy the plateau of the density time-correlator.
A striking example can be found in granular liquids \cite{SH2013}, where
the plateau of the density time-correlator is destroyed by inelastic
collisions.
Hence, we might rather consider that correlations with density-current
modes should be included in general, and it might be accidental that
they can be neglected in sheared thermostated systems.

Note that the result of $\Psi_q(t)$ (which is essentially
$\bar{H}_{\bm{q}}^\lambda(t)$) does not show a growth in the
$\alpha$-relaxation regime, in contrast to $\frac{d}{dt}
\Phi_{\bm{q}}(t)$ (which is essentially $H_{\bm{q}}^\lambda(t)$).
This is an explicit manifestation of the inequivalence of
$H_{\bm{q}}^\lambda(t)$ and $\bar{H}_{\bm{q}}^\lambda(t)$ mentioned previously.

\section{Discussion and Conclusion}

\hspace{0.5em}
At the formal level, response formulas can be derived from the
generaized Green-Kubo (GGK) relation \cite{ME1987, COH2010-3}.
For the case of constant external fields, this relation gives a formula
for the steady-state averages, which characterizes the
relaxation towards a steady state.
In contrast, for the case of time-dependent external fields, the
time-ordering of operators should be handled with care, whose
application to the GGK relation is relatively complicated.
As an alternative, we have constructed an explicit approximate
(quasi-)steady-state distribution function valid on the plaetau of the
density time-correlator, and have derived a response formula at the
linear order.
The situation we consider might be restricted, but at least it is
explicitly calculable, and can be compared to experiments.

In conclusion, we have formulated a nonequilibrium MCT for uniformly
sheared underdamped systems with an isothermal condition.
We have figured out that there is a significant stress relaxation at the
$\alpha$-relaxation regime due to a cooling effect, which is accompanied
by the growth of current fluctuations.
This stress relaxation is predicted to be larger than that in the
overdamped MCT, which is consistent with MD and Brownian dynamics
simulations \cite{ZH2008, AW2013}.
This indicates that nonequilibrium underdamped MCT is not equivalent
to its corresponding overdamped MCT, even at long time-scales after
the early stage of the $\beta$-relaxation.
This is in sharp contrast to the equilibrium case.
We have also shown that correlations with density-current modes are not
important in sheared underdamped thermostated systems, but argued that
this might be accidental, and they should be incoroprated in general.

Another example which cannot be described by the overdamped
formulation is granular particles \cite{SH2013}.
The study of rheological properties of granular liquids by means of
sheared underdamped MCT presented here is in progress, and will be
reported elsewhere \cite{CSOH2012}.

%%%%%%%%%%%%%%%%%%%%%%%%%%%%%%%%%%%%%%%%%%%%%%%%
%% BACKMATTER
%%%%%%%%%%%%%%%%%%%%%%%%%%%%%%%%%%%%%%%%%%%%%%%%

\begin{theacknowledgments}
\hspace{0.5em}
Numerical calculations in this work were carried out at the computer
facilities at the Yukawa Institute and Canon Inc.  
The authors are grateful to S.-H. Chong, M. Otsuki, and M. Fuchs for
stimulating discussions and careful reading of the manuscript, and to
M. Fuchs for provision of the information in Refs.~\cite{ZH2008,
AW2013}.
They also thank Canon Inc. for providing the opportunity of
collaboration.
\end{theacknowledgments}

%%%%%%%%%%%%%%%%%%%%%%%%%%%%%%%%%%%%%%%%%%%%%%%%
%% The bibliography can be prepared using the BibTeX program or
%% manually.
%%
%% The code below assumes that BibTeX is used.  If the bibliography is
%% produced without BibTeX comment out the following lines and see the
%% aipguide.pdf for further information.
%%
%% For your convenience a manually coded example is appended
%% after the \end{document}
%%%%%%%%%%%%%%%%%%%%%%%%%%%%%%%%%%%%%%%%%%%%%%%%

%%%%%%%%%%%%%%%%%%%%%%%%%%%%%%%%%%%%%%%%%%%%%%%%
%% You may have to change the BibTeX style below, depending on your
%% setup or preferences.
%%
%%
%% For The AIP proceedings layouts use either
%%%%%%%%%%%%%%%%%%%%%%%%%%%%%%%%%%%%%%%%%%%%

\bibliographystyle{aipproc}   % if natbib is available
%\bibliographystyle{aipprocl} % if natbib is missing

%%%%%%%%%%%%%%%%%%%%%%%%%%%%%%%%%%%%%%%%%%%
%% You probably want to use your own bibtex database here
%%%%%%%%%%%%%%%%%%%%%%%%%%%%%%%%%%%%%%%%%%%
\bibliography{Suzuki-Hayakawa}

%%%%%%%%%%%%%%%%%%%%%%%%%%%%%%%%%%%%%%%%%%%
%% Just a reminder that you may have to run bibtex
%% All of it up to \end{document} can be removed
%% if you don't like the warning.
%%%%%%%%%%%%%%%%%%%%%%%%%%%%%%%%%%%%%%%%%%%
\IfFileExists{\jobname.bbl}{}
 {\typeout{}
  \typeout{******************************************}
  \typeout{** Please run "bibtex \jobname" to optain}
  \typeout{** the bibliography and then re-run LaTeX}
  \typeout{** twice to fix the references!}
  \typeout{******************************************}
  \typeout{}
 }

\end{document}